\documentclass[aps,prl,reprint,groupedaddress]{revtex4-1}
\usepackage{bm}
\usepackage{graphicx}
\usepackage[usenames]{color}
\usepackage{amsmath,amsfonts,amssymb}
\usepackage{hyperref}
\usepackage{lineno}

\usepackage{ulem}

\begin{document}
\title{Many-Body Chern Number without Integration}

\author{Koji Kudo$^1$}
\author{Haruki Watanabe$^2$}
\author{Toshikaze Kariyado$^3$}
\author{Yasuhiro Hatsugai$^{1,4}$}
\affiliation{
$^1$Graduate School of Pure and Applied Sciences, University of Tsukuba, 
Tsukuba, Ibaraki 305-8571, Japan\\
$^2$Department of Applied Physics, University of Tokyo, Tokyo 113-8656, Japan\\
$^3$International Center for Materials Nanoarchitectonics (WPI-MANA),
National Institute for Materials Science, Tsukuba, Ibaraki 305-0044, Japan\\
$^4$Department of Physics,
University of Tsukuba, Tsukuba, Ibaraki 305-8571, Japan
}

\date{\today}

\begin{abstract}
 The celebrated work of Niu, Thouless, and Wu demonstrated the quantization of Hall conductance in the presence of many-body interactions by revealing the many-body counterpart of the Chern number.
 The generalized Chern number is formulated in terms of the twisted angles of the boundary condition, instead of the single particle momentum, and involves an integration over all possible twisted angles. However, this formulation is physically unnatural, since topological invariants directly related to observables should be defined for each Hamiltonian under a fixed boundary condition. 
 In this work, we show via numerical calculations that the integration is indeed unnecessary -- the integrand itself is effectively quantized and the error decays exponentially with the system size. This implies that the numerical cost in computing the many-body Chern number could, in principle, be significantly reduced as it suffices to compute the Berry connection for a single value of the twisted boundary condition if the system size is sufficiently large.
\end{abstract}

\maketitle

\paragraph{Introduction.}
--- The integer quantum Hall (IQH) states \cite{PhysRevLett.45.494} are the prototypical example of topological phases in condensed matter physics. 
The topological invariant, known as the Chern number or the Thouless-Kohmoto-Nightingale-Nijs (TKNN) invariant, is directly related to the observed quantized Hall conductance~\cite{PhysRevLett.49.405,KOHMOTO1985343}.  The IQH effect demonstrates how topology enriches material phases within the single-particle problem beyond the Landau paradigm where phases are identified solely based on their symmetry breaking patterns. The electron-electron correlation gives birth to even more nontrivial phases with a fractionally quantized Hall conductance~\cite{PhysRevLett.48.1559}.  The fractional quantum Hall (FQH) state for the filling factor $p/q$ exhibits $q$-fold topological degeneracy on torus~\cite{PhysRevLett.55.2095,PhysRevB.40.7387}, which is the defining feature of what we nowadays call ``topological order."

In a pioneering work~\cite{PhysRevB.31.3372}, Niu, Thouless, and Wu developed a many-body generalization of the Chern number [Eq.~\eqref{eq:NTW}], which we call the Niu-Thouless-Wu (NTW) invariant in this work.  The formula equally applies both to the IQH effect in the presence of electron-electron interactions and disorders and to the FQH effect for which interactions are indispensable.  Recently, a wide variety of the strongly correlated topological phases has been identified using the NTW invariant
~\cite{PhysRevLett.90.256802,PhysRevA.76.023613,PhysRevLett.103.105303,
PhysRevB.86.201101,
PhysRevLett.115.116803,PhysRevLett.115.126401,
PhysRevB.95.125134,PhysRevLett.119.177601,
doi:10.7566/JPSJ.86.103701,PhysRevLett.120.096601,Spanton62,
doi:10.7566/JPSJ.87.063701}. 
The formal expression of the TKNN invariant and the NTW invariant are almost identical: The single-particle crystal momentum $\vec{k}=(k_x,k_y)$ and the Bloch wave function in the TKNN invariant are replaced by the twisted angle $\vec{\theta}=(\theta_x,\theta_y)$ of the boundary condition and the many-body ground state in the TKNN integer.  

However, one should keep in mind that at least conceptually there exists a crucial difference between the two formulations. The TKNN invariant is defined for each Hamiltonian under a fixed boundary condition (e.g. the periodic boundary condition).  In contrast, the twisted angle of the boundary condition in the NTW invariant is the integration variable, meaning that the NTW invariant is defined only for a series of the Hamiltonians parametrized by $\vec{\theta}$.  This is unfavorable since topological invariants and the expectation value of physical observables should be computed for each Hamiltonian separately.  
This issue has been recently addressed in mathematical 
works~\cite{Hastings2015,2015arXiv150401243K,2018AIHPA..19..695B,PhysRevB.98.155137}.  It has been getting clear that the integration is indeed 
unnecessary~\cite{PhysRevB.95.121114}
and is no more than a trick making the formula in the same form as the TKNN invariant.  The Berry curvature $F(\vec{\theta})$ [in Eq.~\eqref{eq:sigmatheta} below], computed under a fixed twisted angle, by itself is already ``effectively quantized" in thermodynamically large systems.

The main goal of this work is to confirm the
effective quantization by solving the Hofstadter problems with or
without the electron-electron interaction and disorders. We provide a numerical evidence that the error decays exponentially with
the system size~\cite{Hastings2015,2015arXiv150401243K,2018AIHPA..19..695B,PhysRevB.98.155137}, 
rather than the power-low scaling~\cite{PhysRevB.31.3372}.
The lack of integration not only completes our understanding of the quantized Hall conductance but also may
significantly reduce the computational cost of the Hall conductance, which is
especially advantageous in many-body problems with interactions when the system size is sufficiently large. 
We also discuss the accuracy of the effective quantization (i) in the vicinity of a quantum phase transition and (ii) in the presence of a strong disorder, which are out of the scope of the analytic works~\cite{Hastings2015,2015arXiv150401243K,2018AIHPA..19..695B,PhysRevB.98.155137} that assume a finite excitation gap.

\paragraph{One-plaquette Chern number.}
--- Let us begin by reviewing the relation of the Hall conductance to the NTW invariant. We allow for many-body interactions and impurities, which invalidate the standard description in terms of the Bloch wave function as in the TKNN formalism.  The only assumption is that all excitations are gapped.

We introduce the twisted boundary condition in both the $x$ and $y$ directions and denote the twisted angle by $\vec{\theta}=(\theta_x,\theta_y)$. For a fixed $\vec{\theta}$, the Hall conductance in the linear response theory is given ~\cite{PhysRevB.31.3372}
by $\sigma_{xy}(\vec{\theta})=\frac{1}{q}\frac{e^2}{h}C(\vec{\theta})$, where
\begin{align}
 C(\vec{\theta})
 =- 2\pi iF(\vec{\theta}).
 \label{eq:sigmatheta}
\end{align}
Here,
$F=\frac {\partial  A_y}{\partial \theta _x } -\frac {\partial  A_x}{\partial \theta _y }$
is the Berry curvature associated with the Berry connection $A_\mu= {\rm Tr} \,\Phi^\dagger(\vec{\theta})\frac{\partial \Phi(\vec{\theta})}{\partial\theta_\mu}$ ($\mu=x, y$) and $\Phi(\vec{\theta})=(|G_1(\vec{\theta})\rangle,...,|G_q(\vec{\theta})\rangle)$ is the
ground state multiplet. Note that the integration over $\vec{\theta}$ is absent in Eq.~\eqref{eq:sigmatheta} and the quantization of the Hall conductance is not obvious.

In order to demonstrate the quantization, Ref.~\cite{PhysRevB.31.3372} took the average over all possible values of $\vec{\theta}$, \textit{assuming} that the bulk response has no strong dependence on the boundary condition.
\begin{align}
C=\frac{1}{(2\pi)^2}\int_{T^2}d^2\theta\, C(\vec{\theta})=\frac{1}{2\pi i}\int_{T^2}d^2\theta\,F(\vec{\theta}).
 \label{eq:NTW}
\end{align}
This is the NTW invariant.  Here, $T^2=[0,2\pi]\times[0,2\pi]$ is the torus defined by the twisted angles $\theta_x$ and $\theta_y$. 
In this integrated form, the quantization of the Hall conductance is evident because of the connection to the Chern number, just as in the single-particle problem.

In our following numerical calculation, twisted angles $\theta_\mu$ are discretized into $\theta_\mu=\frac{2\pi}{N_{\theta}}n_\mu$ ($n_\mu=1,2,\ldots,N_{\theta}$). As formulated in Ref.~\onlinecite{doi:10.1143/JPSJ.74.1674}, the discretized Berry curvature $\mathcal{F}(\vec{\theta})
=\log\,
 [U_x(\vec{\theta})
 U_y(\vec{\theta} +\delta_x)
 U_x(\vec{\theta}+\delta_y)^{-1}
 U_y(\vec{\theta})^{-1}]$ is given by the link variables $U_\mu(\vec{\theta})
=\det[\Phi^\dagger(\vec{\theta})
\Phi(\vec{\theta}+\delta_\mu)]
/|\det[\Phi^\dagger(\vec{\theta})
\Phi(\vec{\theta}+\delta_\mu)]|$. Here, $\delta_\mu$'s are defined by $\delta_x=(\frac{2\pi}{N_{\theta}},0)$ and $\delta_y=(0,\frac{2\pi}{N_{\theta}})$.  
The advantage of this formulation is that one does not need to fix the phase of $\Phi(\vec{\theta})$ to make  $\Phi(\vec{\theta})$ a continuous function of $\vec{\theta}$. 
Corresponding to Eqs.~\eqref{eq:sigmatheta} and \eqref{eq:NTW}, we define
\begin{eqnarray}
\mathcal{C}(\vec{\theta})&=&\frac{1}{2\pi i}N_\theta^2\mathcal{F}(\vec{\theta}),\label{eq:opc}\\
\mathcal{C}&=&\frac{1}{2\pi i}\sum_{\vec{\theta}} \mathcal{F}(\vec{\theta})
\label{eq:Chern}.
\end{eqnarray}
We call $\mathcal{C}(\vec{\theta})$ the one-plaquette Chern number.
Even if $N_\theta$ is not so large, $\mathcal{C}$ precisely agrees with $C$~\cite{doi:10.1143/JPSJ.74.1674}, while $\mathcal{C}(\vec{\theta})$ can, in principle, be any real number. We set $N_\theta=20$ for our calculation.
In the Supplemental Material~\cite{sup}, we discuss that our results are
independent of the choice of $N_\theta$.

\paragraph{Noninteracting case.} 
--- Let us start with comparing $\mathcal{C}$ and $\mathcal{C}(\vec{\theta})$ in the noninteracting case. This will set up the stage for our successive discussion on interacting systems. We consider the system of spin-polarized electrons in a uniform magnetic field on a square lattice with $N\times N$ sites. The Hamiltonian with the nearest-neighbor hopping is given by $H_\text{kin}(\vec{\theta})=-t\sum_{\langle ij\rangle}e^{i\phi_{ij}}c_i^\dagger c_j$. 
Here, $c_i^\dagger\ (c_i)$ is the creation (annihilation) operator on site $i$. Because of the twisted boundary condition, they satisfy $c^\dagger_{n_x+N,n_y}=e^{i\theta_x}c^\dagger_{n_x,n_y}$ and $c^\dagger_{n_x,n_y+N}=e^{i\theta_y}c^\dagger_{n_x,n_y}$.  We set $0<t$ in the following. The magnetic field is described by the Peierls phase $\phi_{ij}$, which we fix by the string gauge~\cite{PhysRevLett.83.2246}. The advantage of this gauge choice is that the total number of magnetic flux $N_\phi$ can be freely controlled in the range from $1$ to $N^2$.  The flux per plaquette (the original unit cell) is $\phi=N_\phi/N^2$.  The tight-binding model with $\phi=p/q$ ($p$ and $q$ are co-prime) provides $q$ single-electron bands, where each band has $N^2/q$ single-particle states. Since 
the $p$ low-energy bands form the lowest Landau (LL) level in the weak magnetic 
field limit, the ``LL band'' is given by the set of
these $(N_xN_y/q)\times p= N_\phi$ states.  When the total number of electrons in the system is $N_e$, the filling factor $\nu$ is defined as $\nu=N_e/N_\phi$.  We write the lowest $N_\phi$ eigenvectors of $H_\text{kin}(\vec{\theta})$, belonging to the LL band, as $\psi_k(\vec{\theta})$ ($k=1,2,\ldots,N_\phi$). The creation operator $d_k^\dagger(\vec{\theta})$ of the corresponding state is given by $d_k^\dagger(\vec{\theta})=\bm{c}^\dagger\psi_k(\vec{\theta})$ where $\bm{c}^\dagger=(c_1^\dagger,...,c_{N^2}^\dagger)$.

\begin{figure}[t]
  \begin{center}
   \includegraphics[width=\columnwidth]{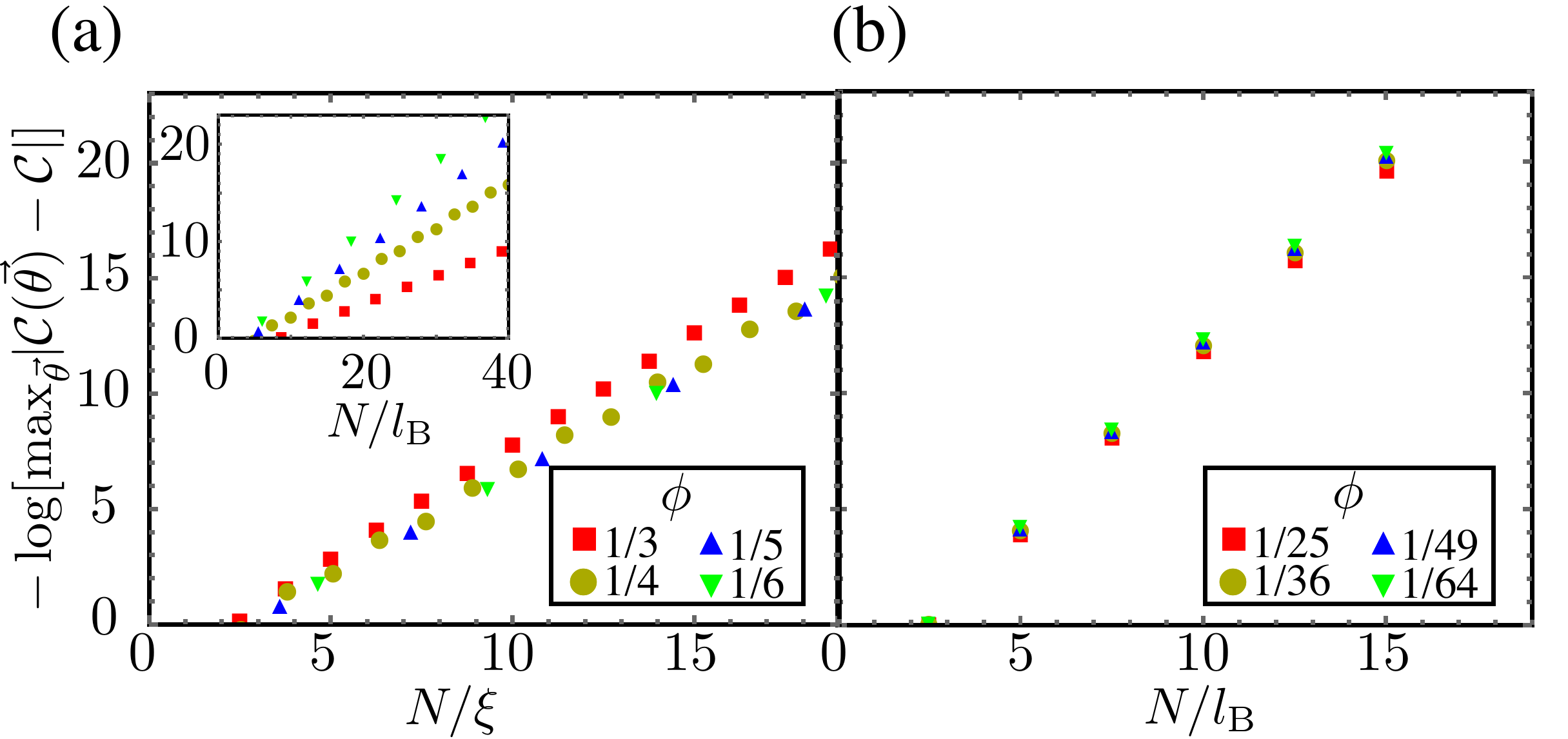}
  \end{center}
 \caption{
 $\max_{\vec{\theta}}|\mathcal{C}(\vec{\theta})-\mathcal{C}|$ as a 
 function of the  system size $N$ scaled by (a) the correlation length $\xi$ or
 (b) the  magnetic  length $l_B$. The inset in (a) shows the 
 $N/l_B$ dependence of $\mathcal{C}(\vec{\theta})$ under a strong magnetic field for
 comparison.}
 \label{fig:Berry_C}
\end{figure}

When the filling factor is $\nu=1$, the ground state is given by completely occupying the LL band as $\Phi(\vec{\theta})=\prod_{k=1}^{N_\phi}d_k^\dagger(\vec{\theta})|0\rangle$ and $\mathcal{C}=1$.  In Fig.~\ref{fig:Berry_C}, we plot $\max_{\vec{\theta}}|\mathcal{C}(\vec{\theta})-\mathcal{C}|$ as a function of the system size $N$ in (a) a strong magnetic field ($\phi\approx1$) and (b) a weak magnetic field ($\phi\ll$ 1).  Here,  `$\max_{\vec{\theta}}$' refers to the maximum value over all $N_\theta^2=400$ plaquettes.  The figure clearly demonstrates that the difference between the one-plaquette Chern number $\mathcal{C}(\vec{\theta})$ and the (averaged) Chern number $\mathcal{C}$ ($=1$ in this case) can be bounded as
\begin{align}
|\mathcal{C}(\vec{\theta})-\mathcal{C}|< A e^{-cN}
 \label{eq:scaling}
\end{align}
with some  coefficients $A$ and $c$. This confirms the validity of using 
$\mathcal{C}(\vec{\theta})$ as the topological invariant for a sufficiently large system
size.

The system size $N$ in Fig.~\ref{fig:Berry_C} is rescaled by the correlation length $\xi$ under a strong magnetic field ($\phi\approx1$) 
and by the magnetic length $l_B=\sqrt{1/(2\pi\phi)}$ under a weak magnetic field ($\phi\ll1$), where $\xi$ is extracted from the correlation function
$\langle c_i^\dagger c_j\rangle$ (see Supplemental Material~\cite{sup} for the details).  Almost all data collapse nicely under this scaling.  This property is lost, for example, when one scales $N$ by $l_B$ in a strong magnetic field regime as shown in the inset in Fig.~\ref{fig:Berry_C} (a).

\begin{figure}[t]
  \begin{center}
   \includegraphics[width=\columnwidth]{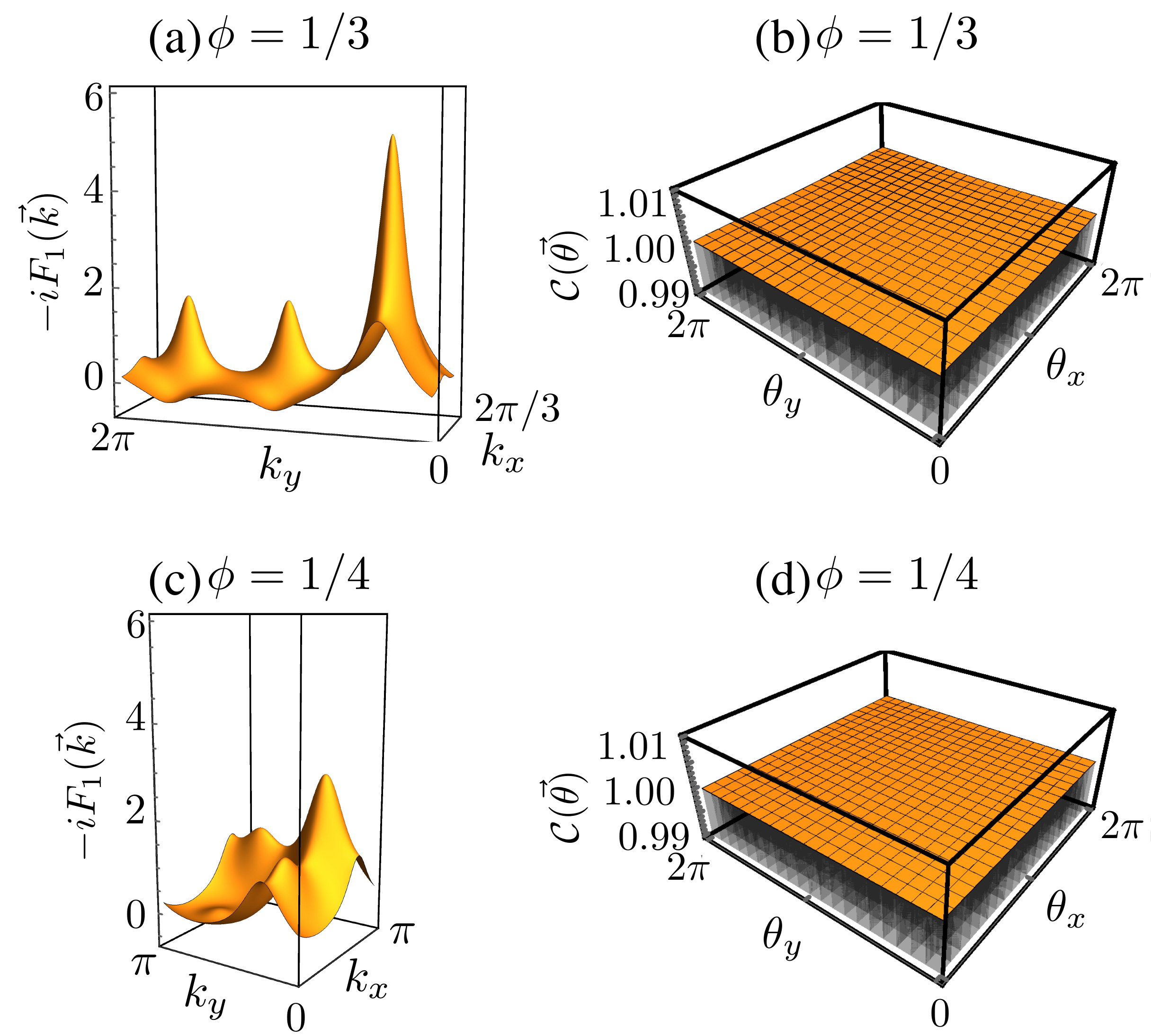}
  \end{center}
 \caption{Single-particle Berry curvature $F_1(\vec{k})$ and 
 one-plaquette Chern number $\mathcal{C}(\vec{\theta})$ at (a),(b) 
 $\phi=1/3$ and (c),(d) $\phi=1/4$.
 The magnetic unit cell is chosen as (a) $3\times1$ and (c) $2\times2$, 
 respectively.  The system size is $N=24$ for (b),(d).}
 \label{fig:1Berry}
\end{figure}

The exponential decay in Eq.~\eqref{eq:scaling} occurs even when the {\it single-particle} Berry curvature $F_1(\vec{k})$ exhibits a strong $\vec{k}$ dependence as far as $F_1(\vec{k})$ is a smooth function of $\vec{k}$. For instance, 
Figs.~\ref{fig:1Berry} (a) and \ref{fig:1Berry} (c) plot $F_1(\vec{k})$ for $\phi=1/3$ and $\phi=1/4$, two values of $\phi$ used in Fig.~\ref{fig:Berry_C} (a), which should be compared to the $\vec{\theta}$ dependence of $\mathcal{C}(\theta)$ in the same setting shown in Figs.~\ref{fig:1Berry} (b) and \ref{fig:1Berry} (d). This result has a simple understanding based on the Euler-Maclaurin formula~\cite{Abramowitz:1974:HMF:1098650}. Let us choose the magnetic unit cells of the size $n_x\times n_y$. Then we have
\begin{eqnarray}
 \mathcal{C}&=&C=\frac{1}{2\pi i}\int_{0}^{\frac{2\pi}{n_x}} dk_x\int_{0}^{\frac{2\pi}{n_y}} dk_yF_1(\vec{k}),\\
 \mathcal{C}(\vec{\theta})&\simeq&C(\vec{\theta})
  =\frac{2\pi }{iN^2}\sum_{i_x=1}^{N/n_x}\sum_{i_y=1}^{N/n_y}
  F_1(\tfrac{2\pi i_x+\theta_x}{N},\tfrac{2\pi i_y+\theta_y}{N}).
  \label{dsum}
\end{eqnarray}
Thus, $C(\vec{\theta})$ approximates the $\vec{k}$ integral in $C$ by a Riemann sum, and $\vec{\theta}$ specifies the choice of representative points from the discretized Brillouin zone.  
Since the Brillouin zone is periodic and $F_1(\vec{k})$ is smooth, the boundary contribution to the correlation exactly vanishes and the error is smaller than any powers of the system size~\cite{PhysRevB.96.085444}.

\paragraph{Interacting case.}
--- We now move on to the interacting problem at the filling factor $\nu<1$.  The electron-electron interactions between the
nearest neighbors can be introduced through the pseudopotential projected onto the LL band~\cite{PhysRevLett.51.605,PhysRevB.86.205424}. The projection matrix $P(\vec{\theta})$ onto the LL band is constructed using the lowest $N_\phi$ eigenvectors $\psi(\vec{\theta})=(\psi_1(\vec{\theta}),...,\psi_{N_\phi}(\vec{\theta}))$ of the kinetic Hamiltonian $H_\text{kin}(\vec{\theta})$ as $P(\vec{\theta})=\psi(\vec{\theta})\psi^\dagger(\vec{\theta})$.  The projected fermion operator is then defined as $\tilde{\bm{c}}^\dagger(\vec{\theta}) =\bm{c}^\dagger P(\vec{\theta})$.  The two-body interaction $V\sum_{\langle ij\rangle}
 c^\dagger_ic^\dagger_jc_jc_i$ ($V$ is the strength of the interaction) can be projected onto the LL band as
$ 
\tilde{H}_\text{int}(\vec{\theta}) =\sum_{klmn}V_{klmn}(\vec{\theta})
 d^\dagger_k(\vec{\theta})d^\dagger_l(\vec{\theta})d_m(\vec{\theta})d_n(\vec{\theta}),
$
where $V_{klmn}=V\sum_{\langle ij\rangle}(\psi_k)_i^*(\psi_l)_j^*(\psi_m)_j
(\psi_n)_i$. Here, the summation over $k,l,m,n$ is 
restricted to the states belonging to the LL band. We choose the interaction strength $V$ in such a way that it is much larger than the band width of the LL band (so that $H_\text{kin}$ can be neglected) but is still much smaller than the cyclotron energy gap (so that higher Landau levels can be neglected). Thanks to this choice, ground states  are obtained by diagonalizing only the interaction Hamiltonian $\tilde{H}_\text{int}(\vec{\theta})$.

With this framework, let us discuss the one-plaquette Chern number $\mathcal{C}(\vec{\theta})$ at the filling factor $\nu=N_e/N_\phi=1/3$.  
Ideally, as we did for the noninteracting case, we would fix the magnetic flux $\phi=N_\phi/N^2$ to a certain value and compute $\max_{\vec{\theta}}|\mathcal{C}(\vec{\theta})-\mathcal{C}|$ as a function of $N$. However, the number of electrons $N_e$ is fixed to be $N_e=\nu N_\phi=(1/3)\phi N^2$ and our current limitation of the numerics is $N_e\leq5$.  For example, when $\phi$ is set to be $1/3$, then $N_e=m^2=1,4,9...$ corresponding to $N=3m=3,6,9,...$.  We would thus get only one data point in the range $2\leq N_e \leq5$, which is certainly insufficient to make figures like Fig.~\ref{fig:Berry_C}.

To overcome this difficulty, we leverage the data collapse for different values of $\phi$ established in the noninteracting limit. We set $N_\phi=N_e/\nu=3N_e$ for $N_e=2$, $3$, $4$, and $5$, and allow $\phi=N_\phi/N^2=3N_e/N^2$ to vary depending on the system size $N$, rather than fix it to a certain number.  Since $N_e\leq 5$, a fairly large system size ($N\sim 20$) naturally falls into the weak magnetic field regime, where the length scale is set by the magnetic length $l_B$. 
\begin{figure}[t!]
  \begin{center}
   \includegraphics[width=\columnwidth]{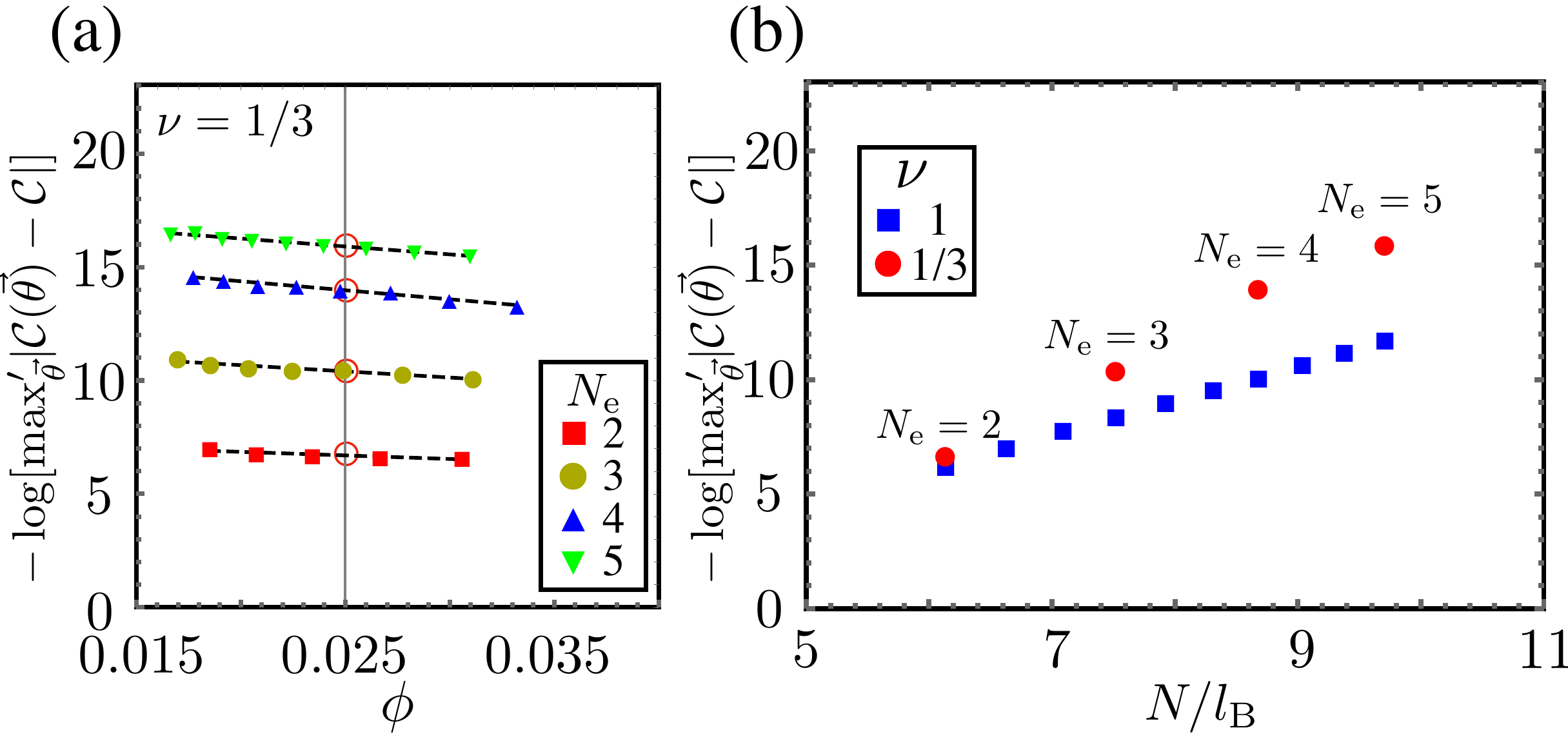}
  \end{center}
 \caption{
 (a) $\max_{\vec{\theta}}'|\mathcal{C}(\vec{\theta})-\mathcal{C}|$ at the filling factor $\nu=1/3$ as a function of the magnetic flux $\phi$ in the range $1/60\leq \phi\leq 1/30$. 
Dashed lines are the linear approximation to find the value for $\phi=0.025$.
 (b) $\max'_{\vec{\theta}}|\mathcal{C}(\vec{\theta})-\mathcal{C}|$ at $\phi=0.025$ for $\nu=1/3$ and $1$. The horizontal axis is the scaled 
 system size $N/l_B=\sqrt{2\pi N_e/\nu}$. }
 \label{fig:MBChern}
\end{figure}

Figure~\ref{fig:MBChern} (a) shows $\max'_{\vec{\theta}}|\mathcal{C}(\vec{\theta})-\mathcal{C}|$ at $\nu=1/3$ as a function of $\phi$, where 
'$\max_{\vec{\theta}}'$' refers to the maximum value over randomly chosen $20$ plaquettes out of in total $N_\theta^2=400$ plaquettes. The system size $N$ is chosen in such a way that $1/60\leq \phi\leq 1/30$. We deduce the value at $\phi=0.025$ for each $N_e$ by a linear extrapolation [see the red circles in Fig.~\ref{fig:MBChern} (a)].  Using these data, we generate Fig.~\ref{fig:MBChern} (b) that plots $\max'_{\vec{\theta}}|\mathcal{C}(\vec{\theta})-\mathcal{C}|$ as a function of the rescaled system size $N/l_B=\sqrt{2\pi N_\phi}=\sqrt{2\pi N_e/\nu}$. 
Although the data point for $N_e=5$ slightly deviates from the 
linear extrapolation from the other system sizes,
Eq.~\eqref{eq:scaling} approximately holds in the $\nu=1/3$ FQH state, implying the exponential accuracy of $\mathcal{C}(\vec{\theta})$ with respect to the system size. As a sanity check of our calculation process, we repeat the same calculation for the noninteracting case with $\nu=1$. The result shown in Fig.~\ref{fig:MBChern} (b) is consistent with the one above.

\paragraph{Quantum phase transition and disordered system.}
--- So far we have investigated the relation between $\mathcal{C}$ and
$\mathcal{C}(\vec{\theta})$ in gapped systems. In the reminder of the work, let us discuss two situations where the excitation gap closes. 

The first example is near a quantum phase transition.  To study this case, let us reuse the noninteracting model above but this time with the next 
nearest-neighbor hopping $-t'$. Here we set $\phi=1/3$ and $\nu=2$. As the value of
$t'/t$ increases, the band cap closes at
$t'/t=2-\sqrt{3}\approx0.268$ and the Chern number jumps from $-1$ to
$+2$~\cite{HATSUGAI2006336}.

\begin{figure}[t]
  \begin{center}
   \includegraphics[width=\columnwidth]{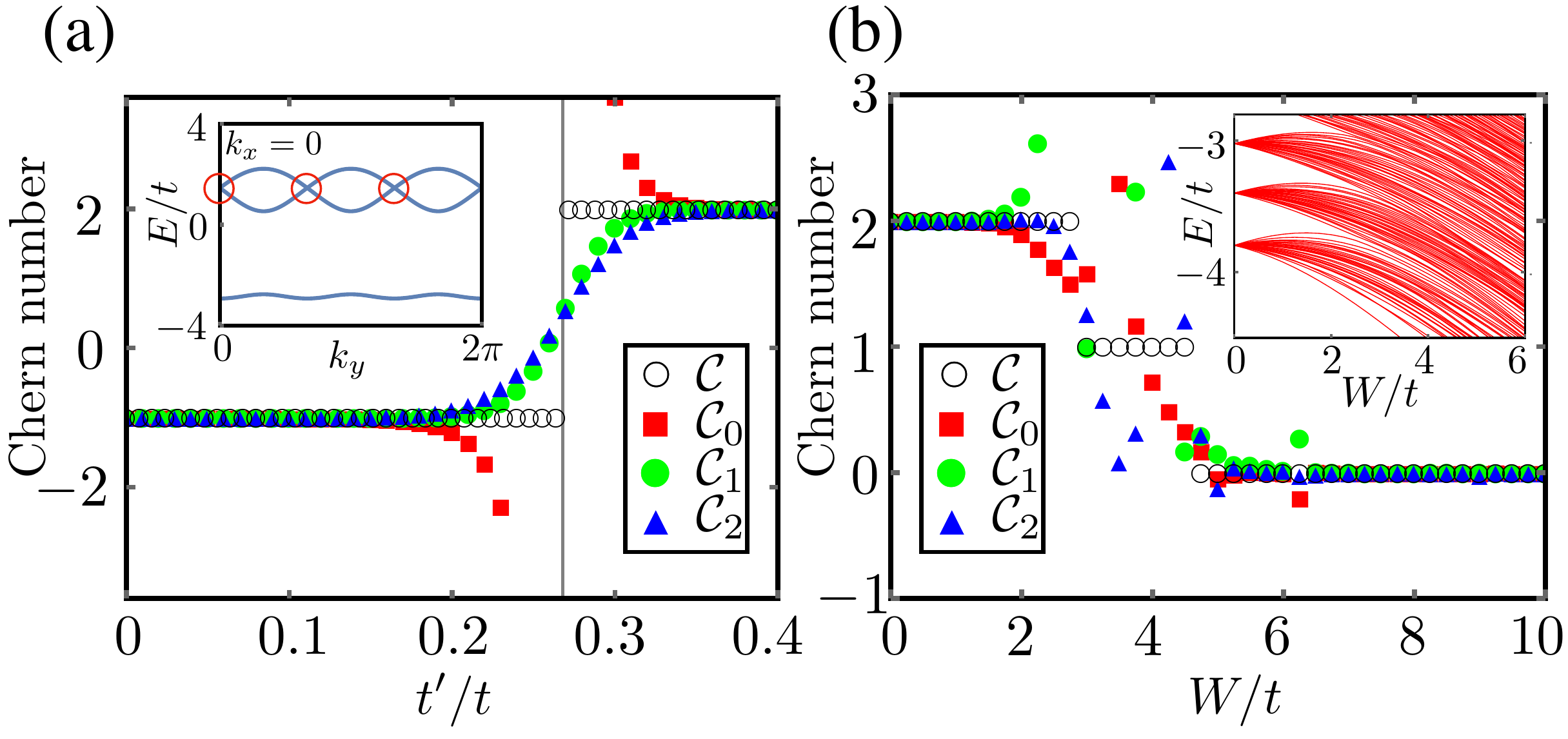}
  \end{center}
 \caption{Chern number $\mathcal{C}$ and one-plaquette Chern numbers
 $\mathcal{C}_0$, $\mathcal{C}_1$ and $\mathcal{C}_2$ of the
 noninteracting model with (a) the next nearest-neighbor hopping ($\nu=2$, 
 $N=30$ and $\phi=1/3$) and (b) the random potential ($\nu=2$, $N=30$ and 
 $\phi=1/30$). The inset in (a) is the energy dispersion at the quantum phase
 transition $t'/t=2-\sqrt{3}$, computed using the magnetic unit cell of
 $3\times1$.  Red circles represent band touching points.
 The inset in (b) is the energy spectrum as a function of the 
 randomness strength $W$.}
 \label{fig:PT}
\end{figure}

In the vicinity of the phase transition, the Berry curvature has sharp peaks at $\vec{k}=(0,0)$, $(0,\frac{2\pi}{3})$, and $(0,\frac{4\pi}{3})$, where the band gap closes as shown in the inset in Fig.~\ref{fig:PT} (a).  Depending on the value of $\vec{\theta}$ and $N$, the discrete summation in Eq.~\eqref{dsum} may or may not contain singular points. Consequently, the one-plaquette Chern number $\mathcal{C}(\vec{\theta})$ in this case is expected to strongly depend on the choice of the plaquettes.  To demonstrate this, here we compute $\mathcal{C}(\vec{\theta})$ for the three choices of plaquettes, $\vec{\theta}=(0,0)$, $(\pi,0)$, and $(\pi,\pi)$. We denote them by $\mathcal{C}_0$, $\mathcal{C}_1$, and $\mathcal{C}_2$, respectively.

Figure~\ref{fig:PT} (a) plots $\mathcal{C}_{0}$, $\mathcal{C}_{1}$, and $\mathcal{C}_{2}$ as a function of $t'/t$.  For $N=30$, the sum in $\mathcal{C}_{0}$ contains all three peaks, while the one for $\mathcal{C}_{1,2}$ includes none of them.  Indeed, we see that only $\mathcal{C}_{0}$ diverges around the transition points. The accuracy of $\mathcal{C}_{1,2}$ also reduces near the transition since the size of the gap becomes smaller.   In general, the correlation length becomes larger as the bulk gap becomes smaller, and the larger system size is required to improve the one-plaquette Chern number.

The second situation is in the presence of a strong disorder.  
We use the original tight-binding model without the next-nearest-neighbor hopping.  When random potentials in the 
range $-W/2$ and $W/2$ are introduced, most of the one-body states become 
localized except the extended states supporting the nonzero Chern number. By increasing
the strength of the random potential, extended states float up in energy across
the Fermi level~\cite{PhysRevLett.52.2304,PhysRevLett.83.2246,PhysRevB.76.132202,PhysRevB.87.115141}. 
As shown in Fig.\ref{fig:PT} (b), the quantum Hall state with 
$\mathcal{C}=2$ becomes the Anderson insulator with vanishing Chern number via 
two successive quantum phase transitions.  We observe that the one-plaquette 
Chern number works well in the two limits: when $W$ is sufficiently small so 
that the excitation gap is fairly big and when $W$ is sufficiently large so 
that states below the Fermi energy are all localized.

\paragraph{Conclusion.} 
--- In this Letter, we demonstrated that the one-plaquette Chern 
number, defined in Eq.~\eqref{eq:opc}, 
is effectively quantized to the true integer value with the exponential accuracy with respect to the linear dimension of the system. 
Our result implies that the one-plaquette Chern number is the legitimate topological number defined for a fixed Hamiltonian characterizing the quantized Hall conductance of a system with many-body interactions and/or disorders assuming that the system is sufficiently large. It also numerically justifies the averaging procedure in the definition of the NTW invariant in Eq.~\eqref{eq:NTW}~\cite{PhysRevB.31.3372}.
Because the one-plaquette Chern number is defined for a fixed boundary condition, it is more physical than the NTW invariant  and it better suits for the experimental situations. 
In addition, the absence of integration is particularly useful in the many-body problems in reducing the computational cost.

\begin{acknowledgments}
 K.K. thanks the Supercomputer Center, the Institute for Solid State 
 Physics, the University of Tokyo for the use of the facilities.
 The work is supported by JSPS KAKENHI Grant Numbers JP17H06138 (K.K., T.K., Y.H.), JP16K13845 (K.K., Y.H.), JP17K17678 (H.W.) and JP17K14358 (T.K.).
\end{acknowledgments}

\bibliographystyle{apsrev4-1}
\bibliography{citation}

\newpage

\renewcommand{\thesection}{S\arabic{section}}
\renewcommand{\theequation}{S\arabic{equation}}
\renewcommand{\thefigure}{S\arabic{figure}}
\renewcommand{\thetable}{S\arabic{table}}
\setcounter{equation}{0}
\setcounter{figure}{0}
\setcounter{table}{0}
\setcounter{page}{1}
\onecolumngrid
\begin{center}
 \LARGE{\bf Supplemental Material}
\end{center}
\vspace{8mm}
\twocolumngrid

\section{$N_\theta$-dependences of one-plaquette Chern number}
\begin{figure}[b]
  \begin{center}
   \includegraphics[width=1.0\columnwidth]{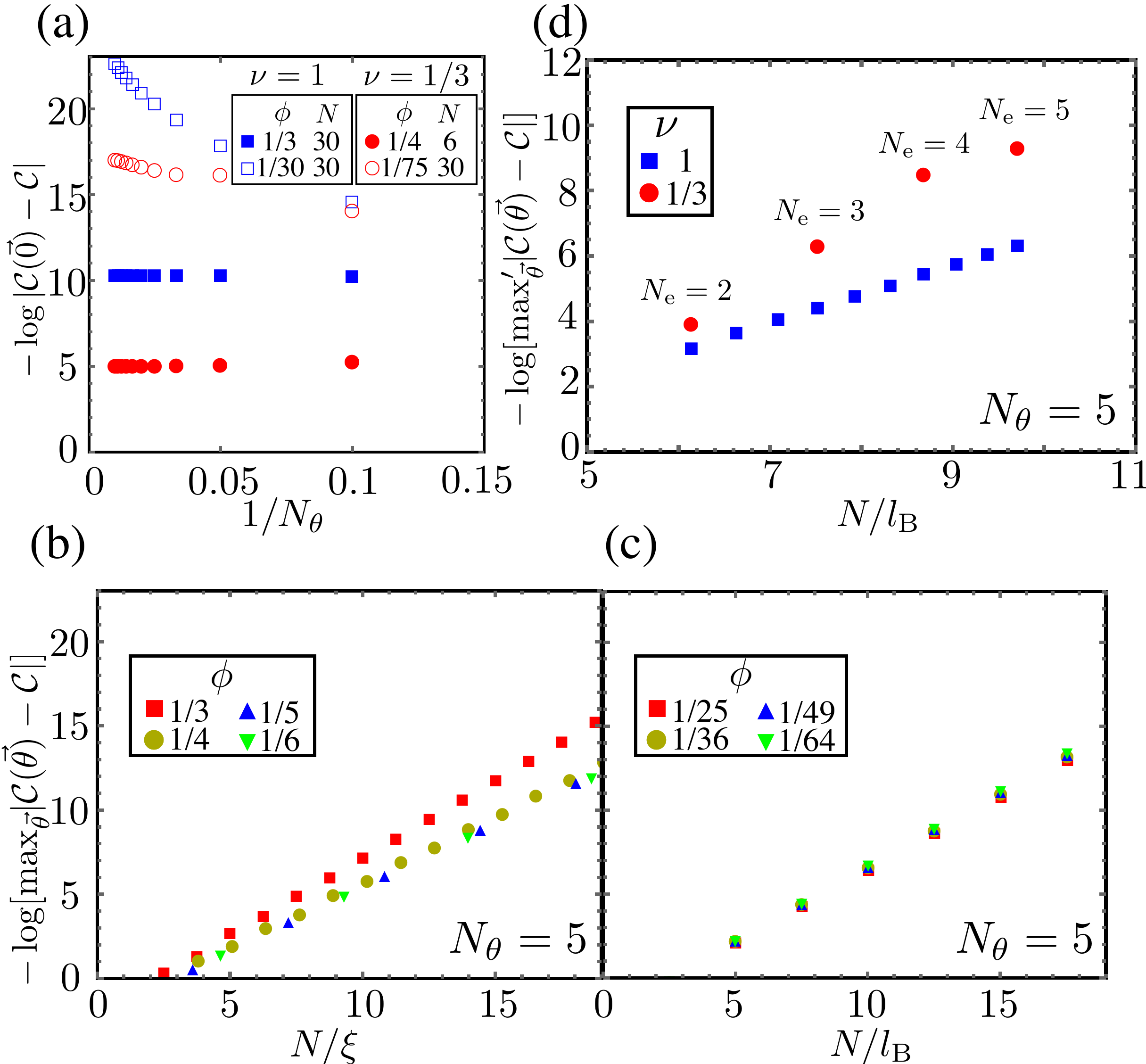}
  \end{center}
 \caption{
 (a) $|\mathcal{C}(\vec{0})-\mathcal{C}|$ as a function of the mesh size $N_\theta$.
 (b,c) The same as Figs.~1 (a,b) in the main text but for $N_\theta=5$.
 (d) The same as Fig.~3 (b) in the main text but for $N_\theta=5$.
}
 \label{fig:OPCN_Ntheta}
\end{figure}

In the main text, the mesh size $N_\theta$ is set as $N_\theta=20$. 
Here we discuss the $N_\theta$-dependence of one-plaquette Chern number $\mathcal{C}(\vec{\theta})$. Figure~\ref{fig:OPCN_Ntheta} (a) displays $\mathcal{C}(\vec{0})$ as a function of 
$N_\theta$ in the non-interacting case at $\nu=1$ and the interacting case at 
$\nu=1/3$. Under a strong magnetic field ($\phi=1/3$ and
$1/4$), $\mathcal{C}(\vec{0})$ is nearly independent of $N_\theta$. However, under a weak magnetic field ($\phi=1/30$ and
$1/75$), $\mathcal{C}(\vec{0})$ becomes more accurate as $N_\theta$ increases.

To double check that the $N_\theta$-dependence does not change our conclusion in the main text, here we include Figs.~\ref{fig:OPCN_Ntheta} (b), (c), and (d) computed under $N_\theta=5$ that correspond to Figs.~1 (a), (b), and Fig.~3 (b) in the main text. They clearly show the identical behavior.

\section{Correlation function in non-interacting system}
\begin{figure}[!t]
  \begin{center}
   \includegraphics[width=1.0\columnwidth]{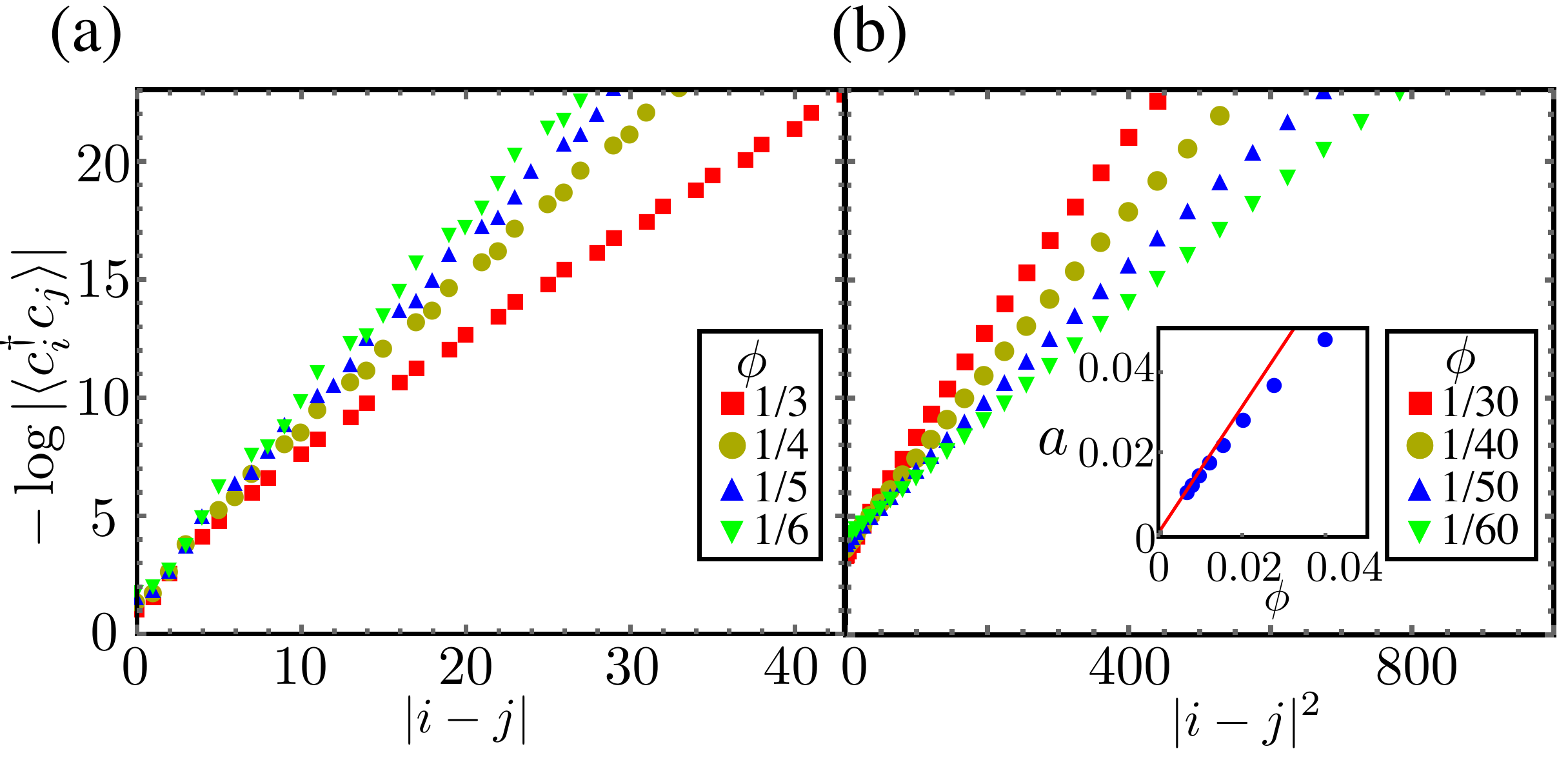}
  \end{center}
 \caption{(a,b)  $|\langle c_i^\dagger c_j\rangle|$ under a magnetic flux (a) $\phi=1/q$ ($q=3$, $4$, $5$, $6$) with $30q\times30q$ sites
 and (b) $\phi=1/q$ ($q=30$, $40$, $50$, $60$) with $3q\times3q$ sites. The 
 inset in (b) shows the exponent $a(\phi)$ under weak magnetic 
 fields. The red line indicates the value $1/(4l_B^2)=\pi\phi/2$ in the continuum limit.
 }
 \label{fig:cc}
\end{figure}

To justify the scaling laws observed in the main text, here we study the behavior of $\langle c_i^\dagger c_j\rangle$ in the non-interacting system. 
In the strong magnetic field  regime, the correlation length $\xi$ is extracted from the correlation function $\langle c_i^\dagger c_j\rangle$. 
Figure~\ref{fig:cc} (a) shows the absolute value of the correlation function for $\phi=1/q$ ($q=3$, $4$, $5$, and $6$) and $N=30q$, which implies the exponential decay of $|\langle c_i^\dagger c_j
\rangle|\propto e^{-|i-j|/\xi(\phi,N)}$. (There are some singular points outside of the range of this plot at $|i-j|=q,2q,\cdots$, where the values of $|\langle c_i^\dagger c_j\rangle|$ are exceptionally small.)  We determine $\xi(\phi,N)$ using the data in the range $0<|i-j|<d_\text{Max}$, where $d_\text{Max}\in[0,N/4]$ is chosen as the maximum value above which $|\langle c_i^\dagger c_j\rangle|$ is always larger than $10^{-10}$. We repeat this process for several $N$'s in the range $100\leq N\leq200$ and find the thermodynamic value by extrapolation. This is what we used in Fig.~1 (a) in the main text.

In the weak magnetic field regime, $|\langle c_i^\dagger c_j\rangle|$ shows a qualitatively different behavior. Figure~\ref{fig:cc} (b) shows the correlation function for $\phi=1/q$ ($q=30$, $40$, $50$, and $60$) and $N=3q$, indicating that
$|\langle c_i^\dagger c_j\rangle|\propto e^{-a(\phi,N)|i-j|^2}$. The constant $a(\phi,N)$ is numerically obtained by the same extrapolation process as we did for $\xi(\phi,N)$. The inset of Fig.~\ref{fig:cc} (b) plots  
$a=\lim_{N\rightarrow\infty}a(\phi,N)$ as a function of $\phi$.  
The behavior in the continuum limit ~\cite{PhysRevLett.58.1252} is known to be $|\langle c^\dagger(z) c(z')\rangle|=(\nu/(2\pi l_B^2))e^{-\frac{1}{4}|z-z'|^2}$ ($z=(x-iy)/l_B$), which is valid for arbitrary filling factor $\nu$ even in the presence of interactions.  The red line in the inset is the value of $a$ of this limit ($1/(4l_B^2)=\pi\phi/2$) and our numerical value of $a$ on the lattice approaches to it in the limit of weak magnetic field, where $l_B$ becomes much larger than the unit cell spacing. These results justify the use of the magnetic length to rescale the system size in this regime.

\end{document}